\documentclass[]{spie}  
\usepackage[dvips]{graphicx}

\title{Humidity contribution to $C_n^2$ over a 600m pathlength in a tropical marine environment}

\author{Mark P. J. L. Chang\supit{1}, Carlos O. Font,
  Charmaine Gilbreath and Eun Oh\supit{2}
\skiplinehalf
\supit{1}Physics Department, University of Puerto Rico, Mayag\"uez, Puerto Rico 00680 \\
\supit{2}U.S. Naval Research Laboratory, Washington D.C. 20375
}

\authorinfo{Further author information: (Send correspondence to M.P.J.L.C.)\\M.P.J.L.C.: E-mail: mchang@uprm.edu, Telephone: 1 787 265 3844}

\begin{document} 
\maketitle 
\begin{abstract}
We present new optical turbulence structure parameter measurements, $C_n^2$,
over sea water between La Parguera and Magueyes Island (17.6N 67W) on the
southwest coast of Puerto Rico.  The 600 meter horizontal paths were located
approximately 1.5 m and 10 m above sea level.  No data of this type has ever
been made available in the literature.  Based on the data, we show that the
$C_n^2$ measurements are about 7 times less compared to equivalent land data.
This strong evidence reinforces our previous
argument\cite{Oh2004a,Oh2004b,Font2006,Chang2006appop} that humidity must be
accounted for to better ascertain the near surface atmospheric turbulence
effects, which current visible / near infrared $C_n^2$ bulk models fail to do.
We also explore the generalised fractal dimension of this littoral data and
compare it to our reference land data.  We find cases that exhibit monofractal
characteristics, that is to say, the effect of rising temperatures during the
daylight hours upon turbulence are counterbalanced by humidity, leading to a
single characteristic scale for the measurements.  In other words, significant
moisture changes in the measurement volume cancels optical turbulence
increases due to temperature rises.
\end{abstract}


\section{INTRODUCTION}
\label{sect:intro}

Bulk climate models estimating and forcasting optical atmospheric turbulence effects are based on the structure parameter (or ``constant'') of temperature
\begin{equation}
\label{eqn:CT2}
C_T^2 = 1.6 \epsilon_\theta \epsilon^{-1/3}
\end{equation}
wherein $\epsilon_\theta$ is the rate of temperature variance dissipation through fluid viscosity (assumed passive additive) and $\epsilon$ is the rate of energy dissipation.  $\epsilon$ is related to the characteristic turbulence mixing length scale $L$ and the energy $\cal{E}$ by the Kolmogorov dimensional approximation
\begin{equation}
\epsilon = 0.7 \frac{{\cal{E}}^{3/2}}{L}
\end{equation}
Gladstone's law then leads to
\begin{equation}
\label{eqn:Gladstone}
C_n^2 = \left( \frac{80 \times 10^{-6} P}{T^2} \right)^2 C_T^2
\end{equation}
The problem with Eq. \ref{eqn:CT2} is the assumption that the temperature variance $\theta$ is a passive additive.  This is false in general since buoyancy forces, which are associated with temperature inhomogeneities, are being ignored.  Neither is there a direct reference to humidity, although that effect is implicit in the definition of $C_T^2$.  To better define the formalism, we have to find an expression for $C_T^2$ in terms of macroscopic variables only.  Masciadri {\em{et al.}}\cite{Masciadri1999} found the following expression for the temperature structure parameter
\begin{equation}
C_T^2 = 0.58 \phi_3 L^{4/3} \left(\frac{\partial \overline{\theta}}{\partial z} \right)^2
\end{equation}
where $\theta$ refers to the virtual potential temperature, $z$ is the
altitude, $L$ is a vertical mixing length for a parcel of air driven by
turbulent forces and $\phi_3$ is a dimensionless number similar to an inverse
Prandtl number.  Obviously, in order to include the macroscopic effects of
buoyancy, moisture and general vertical mixing, $\phi_3$ has to be well
defined.  This is what we seek to do in the marine environment, which
motivates our effort in obtaining $C_n^2$ path integrated measurements in the
visible/near infrared (our instruments' transmitter LEDs are centered on 0.9
$\mu$m, and the detector bandwidth is between 0.65 $\mu$m to 1 $\mu$m) at two altitudes.
 
\section{EXPERIMENT} 
\label{sect:experiment}

\begin{figure}[!htp]
\begin{center}
\begin{tabular}{c}
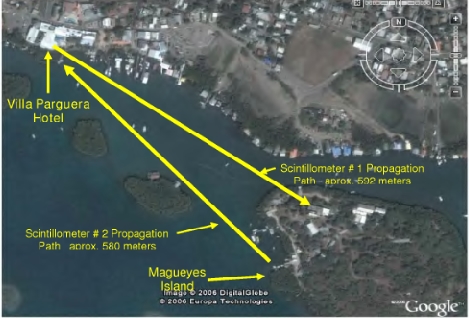
\end{tabular}
\end{center}
\caption{\label{fig:HOPA_VIPh} Location of the littoral environment experiment.}
\end{figure}

\begin{figure}[!htp]
\begin{center}
\begin{tabular}{c}
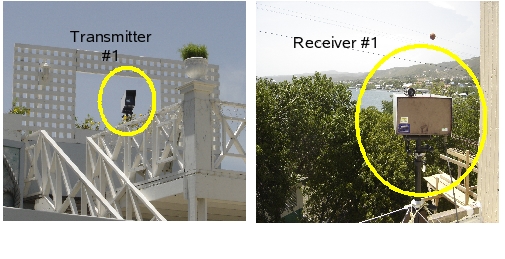 \\
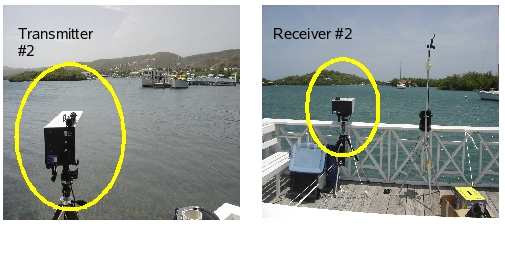
\end{tabular}
\end{center}
\caption{\label{fig:Photos}(Top) Scintillometer 1, defining the path of Beam 1.  The weather station for this level is mounted on the wall above the receiver head.  (Bottom) Scintillometer 2, defining the path of Beam 2.  The other weather station can also be seen next to the receiver head.}
\end{figure}

The objective of the experimental configuration, shown in Figure \ref{fig:HOPA_VIPh} was to provide a comparison of the path integrated $C_n^2$ at two different altitudes within the surface layer over water with data previously recorded over a land bound site\cite{Chang2006,Font2006MS,Chang2006appop} (between February and April 2006, the run is designated ``$\Phi^r$''); all of the measurements were carried out along the west coast of Puerto Rico (sited approximately at 18N 67W).  
Two OSI LOA-004 scintillometers\cite{Font2006} and two Davis Provantage Plus (DP+) weather stations were sited between the University of Puerto Rico's Magueyes Island Research facility and a private site on the coastal section of the town of La Parguera, Lajas.  For ease of reference, this campaign is designated HOPA VIPh (HOrizontal PAth propagation from the VIlla Parguera hotel).  
As illustrated, two beam paths were defined: scintillometer 1 (beam 1) was located between the rooftops of buildings, such that approximately 14\% of the optical path length crossed over land, while scintillometer 2 (beam 2) was configured to record a pure over water path between jettys.  Both scintillometer sample rates were set to 0.1Hz. 
The altitude of beam 1 varied from 6 to 15 m end--to--end, with a path length of 580 m, while that of beam 2 was a constant 1.5 m above the sea surface, with a path length of 592 m.   The DP+ weather stations were closely located with the receiver heads of each instrument.  Figure \ref{fig:Photos} show photographs of the instrument installations.

The experimental campaign was carried out over 3 weeks between 20 August and 5 September of 2006, corresponding to year days 232 through to 248.

\section{DISCUSSION OF DATA}
\label{sect:data}

We obtained 14 datasets (spanning a full 24 hour period) per scintillometer.  
Beam 1, at a mean height of 10 m above sea level, provided 5 sets with 95\% or
higher valid data.  In the case of the near sea--surface beam 2, only 3 sets were totally free from data dropouts, mostly arising from wind misalignments.  It is probable that human activity (primarily boats near the beam) affected the full range of records, given that the beam path crossed a public marine space.  These show themselves as strong data spikes, but in the absence of independent confirmation for each event we have not attempted to filter them. 
\begin{figure}[!htp]
\begin{center}
\begin{tabular}{c}
HOPA\_VIPh\_2altitudes.jpg
\end{tabular}
\end{center}
\caption{\label{fig:HOPA_VIPh_2altitudes} The complete record of 24 hour data,
  smoothed with a forward boxcar average of 60 points (5 minutes).  The dots
  are from beam 1 (at 10 m mean altitude), and the crosses represent beam 2
  (at 1.5 m altitude).  The ordinates run from $10^{-9}$ to $10^{-10}$
  m$^{-2/3}$.}
\end{figure}
In figure \ref{fig:HOPA_VIPh_2altitudes} we present the 14 complete 24 hour datasets.  Note that while the two beams do track each other quite well, the lower altitude measurements show more spread around the daylight hours than the higher altitude.  We attribute this to human activity since there does not seem to a regular overall pattern.  Most notable in the data is the partial or complete suppression of the daytime `D' curve in the $C_n^2$ data. 

By way of comparison, in figure \ref{fig:HOPA_PhysRoof_both}, we present a
subset of the $\Phi^r$ landbound record measured with the same instruments.
The $\Phi^r$ configuration differed in path length from that of HOPA VIPh; it
was much shorter, at 90 m between transmitter and receiver.  In this run, the
scintillometers were placed side by side, at the same altitude.  While we do not claim that it is valid to undertake a direct comparison of all the moments of the landbound ensemble with HOPA VIPh, it is certainly reasonable to examine the difference between means.  The mean value for $\Phi^r$'s path integrated $C_n^2$ is approximately 6.6 times larger than that of HOPA VIPh. 
\begin{figure}[!htp]
\begin{center}
\begin{tabular}{c}
HOPA\_PhysRoof\_both.jpg
\end{tabular}
\end{center}
\caption{\label{fig:HOPA_PhysRoof_both} Landbound $C_n^2$ record from the
  $\Phi^r$ run.  Dots refer to scintillometer 1 and crosses to
  scintillometer 2, as defined for HOPA VIPh.  The instruments were adjacent
  to each other at the same altitude, with a distance of 90 m between
  transmitter and receiver.  The ordinates are at the same scale as Figure
  \ref{fig:HOPA_VIPh_2altitudes}.}
\end{figure}

\section{FRACTAL ANALYSIS}

In order for the $C_n^2$ measurements presented in Section \ref{sect:data} to be easily compared to future data under different conditions, we have begun a preliminary study of the dimensional (fractal) structure of the temporal distribution.  The idea that optical turbulence potentially has a fractal structure is not new.  In fact Schwartz {\em{et al.}} identified that Kolmogorov turbulence degraded wavefronts resemble a fractional Brownian motion fractal surface \cite{Schwartz1994}.

Determination of fractal dimensions from experimental data is not a straightforward process, due to the many sources of error introduced by the finiteness of datasets involved\cite{Grassberger1988} and the finite range of length scales inherent therein.  Borgani {\em{et al.}} \cite{Borgani1993} surveyed the most popular methods for determining fractal dimensions: generalized correlation integrals, box counting methods, density reconstruction procedures and minimal spanning tree algorithms.  Their conclusion is that nearest neighbour and minimal spanning tree methods fail to accurately gauge the fractal dimensions, principally because they discard information about the inter--point distance.  Of the remaining three options, box counting was found to be very similar to correlation integrals; since box counting is anisotropic (therefore very slow), we discard them from our consideration here.  

It turns out that correlation integrals are most reliable for generalised fractal dimensions, $D_q$, of positive order, and less reliable for negative order.  In contrast, the density reconstruction procedure is reliable for negative order and less so for positive order.  Negative order fractal dimensions measure the ``emptiness'' of a distribution, while positive order dimensions measure the space (or time in our case) filling of a distribution.
We are interested in the measure of (time) filling, so we have made use of the correlation integral procedure, due to Pawelzik and Schuster\cite{Pawelzik1987}.

\subsection{CORRELATION INTEGRAL ESTIMATES OF FRACTAL DIMENSION}

Before detailing the results of our preliminary findings, we review the ideas behind the correlation integral.  Given a finite set of $N$ data points $x_i$ sampled from some distribution, the correlation functions $C_i(t)$ are defined to be the fraction of points within time interval $t$ of the $i$th data point.  The partition function
\begin{equation}
Z(t,q) = \frac{1}{N} \sum^N_{i=1} C_i(t)^{q-1}
\end{equation}
is then expected to scale as $r^\tau$ over the range of length scales where fractal (self similar) properties hold.  If we fit a power law to $Z(t,q)$ then we can find the generalised fractal dimension $D_q$ through
\begin{equation}
\tau = (q-1)D_q
\end{equation}
For example, the Hausdorff dimension is estimated if $q=0$, the information dimension is estimated if $q\rightarrow 1$ and the correlation dimension is from $q = 2$.  We note here that the estimation of the $q=0$ dimension is extremely sensitive to regions of low point density (data dropouts) so although we list it in what follows, it is probably unwise to regard it as a ``safe'' universal measure.
\begin{table}[!htp]
\begin{center}
\begin{tabular}{llllllll} \hline
Date    & Beam  & \% Data       & \%    & $D_0$ & $D_1$ & $D_2$ & $D_4$   \\ 
        &       & Recorded      & Valid &       &       &       &         \\ \hline
08/22/06& 1     & 99.1 & 98.38 & 0.83(12) & 0.90(18) & 0.91(36) & 0.94(22) \\
08/23/06& 1     & 100  & 99.98 & 0.80(22) & 0.90(52) & 0.93(04) & 0.97(69) \\
08/24/06& 1     & 95.6 & 96.15 & 1.05(18) & 0.94(74) & 0.98(48) & 0.88(57) \\
08/29/06& 1     & 100  & 100   & 0.96(68) & 0.95(22) & 0.97(70) & 1.04(41) \\ 
08/22/06& 2     & 100  & 96.75 & 0.96(89) & 0.89(65) & 0.91(41) & 0.88(82) \\
08/23/06& 2     & 100  & 95.76 & 0.95(65) & 0.90(40) & 0.87(51) & 0.94(24) \\
08/24/06& 2     & 100  & 88.59 & 0.99(92) & 0.99(24) & 0.97(83) & 0.97(62) \\
08/25/06& 2     & 100  & 99.41 & 0.96(05) & 0.93(89) & 0.96(89) & 0.95(72) \\
{\em{03/09/06}}& - & {\em{100}} & {\em{100}} & {\em{0.97(21)}}  &  {\em{0.95(08)}}  & {\em{0.85(36)}}  & {\em{0.90(87)}}  \\ \hline
reference fractal & & (64 points) & & 0.99(84) & 0.98(63) & 0.98(45) & 1.00(88) \\ \hline
\end{tabular}
\caption{\label{tbl:Dq} Generalised fractal dimensions of HOPA VIPh datasets.
        This is compared with an equivalent $\Phi^r$ run, in italics.  The
        dimension were calculated for data that had been smoothed with a 60
        point forward moving boxcar. Beam 1 is the high altitude beam, beam 2
        is the low altitude beam. On the last line of the table, we present
        the estimated generalised dimensions for a 64 point theoretical pure
        fractal of dimension 1.0000.}
\end{center}
\end{table}
From the data shown in Table \ref{tbl:Dq}, we can discern the accuracy of the correlation integral algorithm from the results of applying it to a reference 64 point fractal of dimension 1.0000.  $D_{0,1,2,4}$ lie no more than 1.55\% from the true dimensional value, and their mean varies from 1.0000 by 0.6\%.  The $C_n^2$ data series themselves have 8640 points (assuming no data gaps), so we anticipate that the estimation accuracy will be no worse than the reference values.

If we examine the variance from the dimensional mean of each of the datasets shown, the 08/24/06 and 08/25/06 sets from the low altitude beam exhibit monofractal characteristics (mean dimensions of 0.99 and 0.96 respectively, with differences with respect to the mean of no more than 2\%), while the rest are multifractal.  We expect that fully developed turbulence will be concentrated on a multifractal set: the fact that we see a possible monofractal set implies that those data possess a global dilatation invariance, regardless of solar insolation.

We claim that injection of moisture into the measurement volume for optical turbulence dampens out the temperature effect; this result occurs for the 1.5 m altitude beam but from our preliminary calculations, seems not to occur in the 10 m beam.

\section{CONCLUSIONS}

We have presented $C_n^2$ data acquired at two altitudes, 1.5 m and approximately 10 m, above the sea surface in the tropical littoral environment of La Parguera, Lajas, Puerto Rico over a 0.6 km pathlength.  In comparison to 90 m landbound measurements, we note the suppresion of increased turbulent activity due to elevated temperatures when the sun is up.  We also found that the mean value of $C_n^2$ is some 6.6 times less for the marine measurement ensemble compared to the landbound readings, strongly indicating that moisture plays an equivalently strong role to temperature in determining $C_n^2$.  Finally we have presented initial estimates for the general fractal dimensions $D_q$, obtained with the correlation integral method due to Pawelzik and Schuster.  This shows that the data taken very close to the surface of the sea can behave as a monofractal, therefore indicating that humidity plays a very significant part in countering the effect of rising temperature during the daylight hours upon optical turbulence.  It may be possible to formulate this in terms of a detailed balance (in wavenumber space) between input and output energy at only one scale for these cases.

\acknowledgments     
Part of this work was sponsored by the Office of Naval Research.


\bibliography{bib}   
\bibliographystyle{spiebib}   

\end{document}